\documentclass[a4paper]{jpconf}

\usepackage[sumlimits,intlimits,namelimits]{amsmath}
\usepackage{amssymb}

\usepackage{graphicx}

\newcommand{\mA}{\mathcal{A}}

\newcommand{\be}{\begin{eqnarray}}
\newcommand{\ee}{\end{eqnarray}}

\def\>{\rangle}
\def\<{\langle}

\usepackage[bordercolor=black,backgroundcolor=yellow,linecolor=black,
textsize=scriptsize,shadow]{todonotes}

\begin{document}
\title{Time dependence of entanglement for steady state formation in AdS$_3$/CFT$_2$}

\author{M Flory${}^{1}$, J Erdmenger${}^{2}$, D Fern\'andez${}^{3}$, E Meg\'ias${}^{4}$, A Straub${}^{5}$ and P Witkowski${}^{6}$}

\address{${}^{1}$ Institute of Physics, Jagiellonian University, \L{}ojasiewicza 11, 30-348 Krak\'ow, Poland}
\address{${}^{2}$ Institut f\"ur Theoretische Physik und Astrophysik, Julius-Maximilians-Universit\"at W\"urzburg, Am Hubland, 97074 W\"urzburg, Germany}
\address{${}^{3}$ University of Iceland, Science Institute, Dunhaga 3, 107 Reykjav\'ik, Iceland}
\address{${}^{4}$ Departamento de F\'{\i}sica Te\'orica, Universidad del Pa\'{\i}s Vasco UPV/EHU,
	Apartado 644,  48080 Bilbao, Spain}
\address{${}^{5}$ Max-Planck-Institut f\"ur Physik (Werner-Heisenberg-Institut),
	F\"{o}hringer Ring 6, 80805 M\"unchen, Germany}
\address{${}^{6}$ Max-Planck-Institut f\"ur Physik komplexer Systeme,
	N\"othnitzer Stra{\ss}e 38, D-01187 Dresden, Germany}

\ead{mflory@th.if.uj.edu.pl}

\begin{abstract}
We consider a holographic model of two 1+1-dimensional heat baths at different temperatures joined at time $t=0$, such that a steady state heat-current region forms and expands in space for times $t>0$. After commenting on the causal structure of the dual 2+1-dimensional spacetime, we present how to calculate the time-dependent entanglement entropy of the boundary system holographically. We observe that the increase rate of the entanglement entropy satisfies certain bounds known from the literature on entanglement tsunamis. Furthermore, we check the validity of several non-trivial entanglement inequalities in this dynamic system.
\end{abstract}
\section{Introduction and setup}
\label{sec::intro}

During the recent years, there have been a number of fruitful attempts to apply methods of gauge/gravity duality to the study of strongly coupled far-from-equilibrium systems. In these proceedings, following our recent paper \cite{Erdmenger:2017gdk}, we will investigate the specific holographic system studied in \cite{Bhaseen:2013ypa}. There, a setup was proposed where at time $t=0$, two semi-infinite heat reservoirs are brought into contact, leading to an initial temperature profile of the form  
\begin{align}
T(t=0,x)= T_\mathrm{L}\,\theta(-x)+T_\mathrm{R}\,\theta(x),
\label{introduction: initial temperature profile}
\end{align}
where $x$ is a spacial coordinate. In what follows, we will focus on the case where the field theory is $1+1$ dimensional, leading to a $2+1$ dimensional holographic bulk description. 

\begin{figure}[h]
	\begin{center}
	\begin{tabular}{ccc}
		\hspace{-0.4cm}\includegraphics[width=0.3\linewidth]{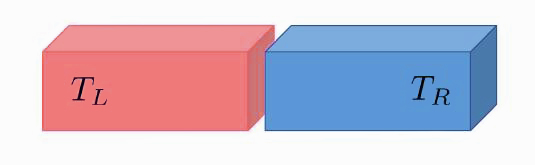} &
		\hspace{-0.7cm}\includegraphics[width=0.1\linewidth]{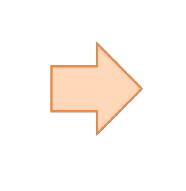} &
		\hspace{-0.5cm}\includegraphics[width=0.3\linewidth]{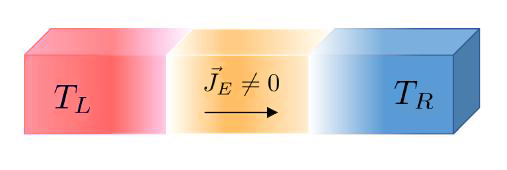} \\
	\end{tabular}
	\caption{At $t=0$, the temperature profile is given by equation \eqref{introduction: initial temperature profile}. Evolving forward in time from $t=0$, a spatially homogeneous non-equilibrium steady state develops in the middle region, carrying an energy current $J_E$.}
	\label{fig:system}
	\end{center}
\end{figure}

On the field theory side, it can be shown that as time evolves onward from the initial condition \eqref{introduction: initial temperature profile}, a constant energy flow $\left\langle J_E\right\rangle\neq0$, the \textit{steady state}, develops in a growing region between two shockwaves emerging at the speed of light from $t=x=0$ \cite{Bhaseen:2013ypa}, see figure \ref{fig:system}. In contrast to similar setups in higher dimensions, this $1+1$ dimensional setup has the special property that the shock waves with which the steady state region expands are dissipation-free, meaning that the temperature profile evolving from the initial conditions \eqref{introduction: initial temperature profile} will involve sharp step-functions even for times $t\gg0$.

\begin{figure}[htb]
	\includegraphics[width=0.43\textwidth]{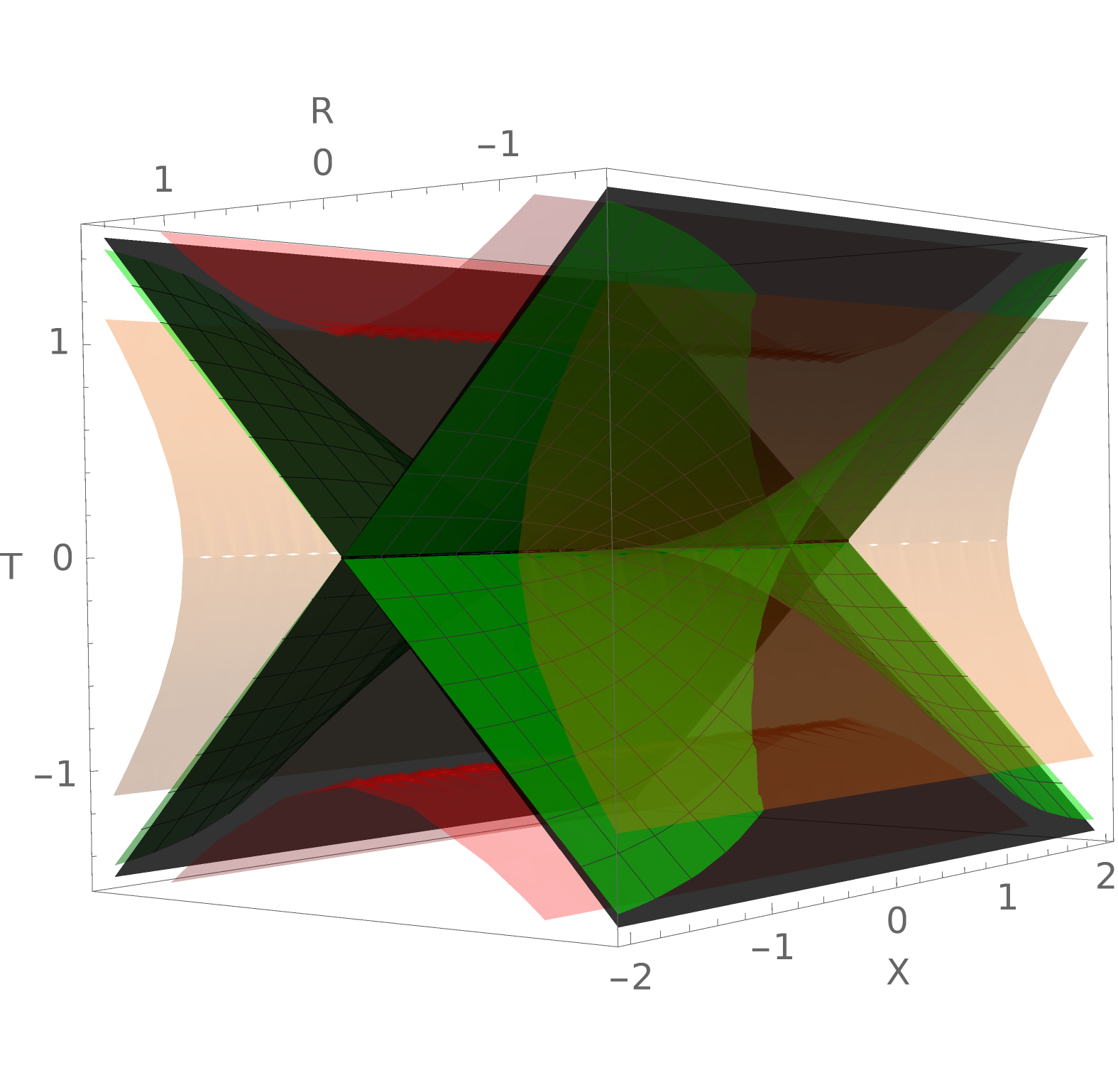}\hspace{0.07\textwidth}%
	\begin{minipage}[b]{0.5\textwidth}	\caption{Three dimensional Kruskal diagram (with Kruskal coordinates $T,R,X$, see \cite{Erdmenger:2017gdk}) of the bulk spacetime \eqref{eq:FG}. The black diagonal sheets correspond to the event horizons. The red surfaces show the singularities, and the orange surfaces show the location of the boundary at $z=0$. The green surface is the worldvolume of the bulk hypersurfaces along which the different regions of spacetime are glued together. A steady state region forms both for $t>0$ and for $t<0$. Only the $t>0$ region is considered to be physically relevant.}
		\label{graf:Kruskal}
	\end{minipage}
\end{figure}

Our goal will be to use holographic methods \cite{Ryu:2006bv,Ryu:2006ef,Hubeny:2007xt} in order to calculate the time dependence of entanglement entropy in this system. The holographic dual to the $1+1$ dimensional large $c$ CFT described above is given by an asymptotically AdS$_3$ (vacuum) solution to Einsteins equations which generically takes the form \cite{Banados:1998gg,Bhaseen:2013ypa}
\begin{subequations}
	\begin{align}
ds^2&=\frac{1}{z^2}\left( d z^2 +  g_{\mu\nu}(z, x,t) d x^\mu d x^\nu  \right) \,, \label{eq:FG}
\\
	g_{tt}(z,x,t) &= -\left[1-z^2\left( f_R(x-t) + f_L(x+t) \right)\right]^2  +  \left[ z^2\left( f_R(x-t) - f_L(x+t) \right) \right]^2  \,,  \\
	g_{tx}(z,x,t) &=   -2 z^2\left( f_R(x-t) - f_L(x+t) \right)  \,,  \\
	g_{xx}(z,x,t) &= \left[1+z^2\left( f_R(x-t) + f_L(x+t) \right)\right]^2  -  \left[ z^2\left( f_R(x-t) - f_L(x+t) \right) \right]^2  \,, 
	\end{align}
	\label{eq:gxx}
\end{subequations} 
where we have set the AdS-radius $L\equiv1$ and $f_R(v)$ and $f_L(v)$ are arbitrary functions to be determined from the initial conditions. Specifically, demanding \eqref{introduction: initial temperature profile} to hold in the dual CFT enforces the choice \cite{Bhaseen:2013ypa}
\begin{equation}
f_{L/R}(v) \equiv \frac{\pi^2 }{2}\left( T_L^2 + \left( T_R^2 - T_L^2 \right)\theta(v) \right) \,. \label{eq:fLfRlimit}
\end{equation}
The appearance of the step function in \eqref{eq:fLfRlimit} means that the bulk spacetime will appear to consist of several sectors describing static and boosted BTZ black holes, matched together along co-dimension one hypersurfaces. This was discussed in more details in \cite{Erdmenger:2017gdk}, see also figure \ref{graf:Kruskal}. The important thing to notice is that these matching hypersurfaces are spacelike in nature, hence they cannot be interpreted as physical objects carrying information in the bulk. 

\section{Entanglement entropy}
\label{sec::EE}

For static bulk spacetimes, the entanglement entropy $S(A)\equiv -\text{Tr}\left[\rho_A \log(\rho_A)\right]$ for a boundary subregion $A$ with reduced density matrix $\rho_A$ can be calculated via the Ryu-Takayanagi (RT) proposal \cite{Ryu:2006bv,Ryu:2006ef}
$
S_A(A)=\frac{\mA(\gamma_A)}{4 G_N},
$ 
where $\mA(\gamma_A)$ is the area of a minimal spacelike codimension two bulk surface $\gamma_A$ anchored on the asymptotic boundary. The generalisation of this prescription to time-dependent bulk spacetimes was discussed in \cite{Hubeny:2007xt}. In this Hubeny-Rangamani-Takayanagi (HRT) proposal, the bulk surface $\gamma_A$ is merely required to be an \textit{extremal} hypersurface. In the $2+1$ dimensional bulk setup \eqref{eq:FG} of interest here, the curves $\gamma_A$ will be spacelike geodesics.

As discussed in \cite{Erdmenger:2017gdk}, there are two possibilities to tackle the geodesic problem in the background \eqref{eq:FG}. The seemingly most direct way would be to solve the geodesic equations numerically. Due to the way in which the geodesics have to be anchored at the boundary, this would require the use of non-trivial numerical techniques such as \textit{relaxation methods} (similar to \cite{Ecker:2015kna}) or \textit{shooting methods} as successfully implemented for the system of our interest in \cite{Megias:2015tva,Megias:2016vae,Erdmenger:2017gdk}.  

Due to the piecewise structure of \eqref{eq:FG} discussed in section \ref{sec::intro}, there is however a more elegant way to solve this problem in the system of interest: The spacetime can be divided into four sectors $(t+x>0 \wedge x-t>0), (t+x>0 \wedge x-t<0),(t+x<0 \wedge x-t>0)$ and $(t+x<0 \wedge x-t<0)$, see figure \ref{graf:Kruskal}. As in each of these quarters the bulk metric is the one of a (boosted) BTZ black hole \cite{Banados:1992wn}, and as geodesic curves and distances between spacelike separated points in a BTZ spacetime are known analytically \cite{Shenker:2013pqa}, it is possible to write the (renormalised) geodesic distance between two given boundary points of the spacetime \eqref{eq:FG} as a function of the coordinates of the matching points where the corresponding geodesic passes through the hypersurfaces along which the different sectors are glued onto each other. Extremising the resulting length with respect to the coordinates of the matching points yields then the correct overall geodesic length. This can be seen as a way to derive the appropriate refraction conditions for geodesics at the hypersurfaces. For situations where the geodesic in question crosses the hypersurface only once, this procedure has been worked out in detail in \cite{Erdmenger:2017gdk}. It was found that the relevant equations can be solved analytically in the cases where either both $T_R$ and $T_L$ are perturbatively small or where $T_R=0$ exactly with arbitrary finite $T_L$. Otherwise, the numerical problem boils down to finding the solutions to a set of \textit{algebraic} equations.

\section{Results}
\label{sec::Results}

\subsection{General behaviour}
\label{sec::General}

\begin{figure}[htb]
	\begin{minipage}{0.475\textwidth}
		\includegraphics[width=\textwidth]{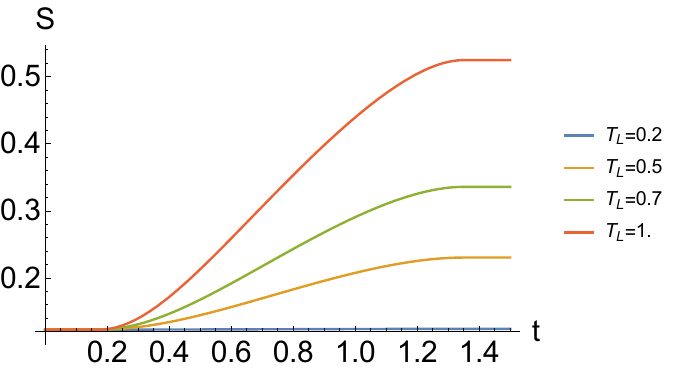}
		\caption{Renormalised entanglement entropies $S_A(t)$ for a boundary interval $0.175\leq x\leq1.35$, $T_R=0.195$ and $T_L$ ranging from $0.2$ (lowest curve) to $1.0$ (highest curve) as a function of boundary time $t$.}
		\label{fig:EntanglementEntropies}
	\end{minipage}\hspace{0.05\textwidth}%
	\begin{minipage}{0.475\textwidth}
		\includegraphics[width=\textwidth]{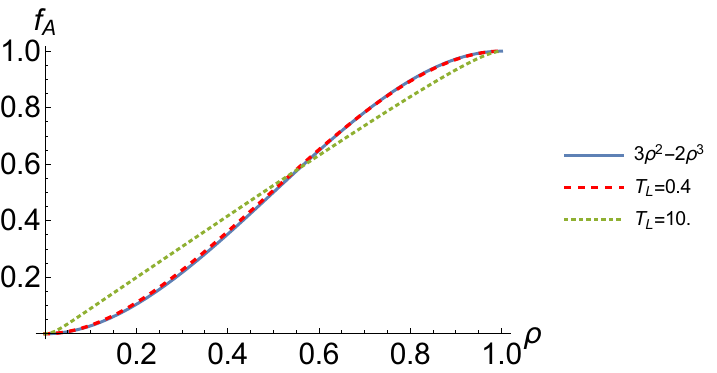}
			\caption{Rescaled entanglement entropy $f_A(\rho)$ for the interval of figure \ref{fig:EntanglementEntropies} with $T_L=0.4$ and $T_L=10$ compared to the universal formula \eqref{eq:fAuniversal}. }
			\label{fig:UniversalFormula}
	\end{minipage} 
\end{figure}

Figures \ref{fig:EntanglementEntropies} and \ref{fig:UniversalFormula} show representative results obtained in \cite{Erdmenger:2017gdk} with the method described in section \ref{sec::EE}. There are three important observations to be made: Firstly, the time evolution of entanglement entropy is \textit{causal}, i.e.~before the shockwave moving over the boundary with the speed of light from $t=x=0$ enters the interval and after it has left the interval, the entanglement entropy of a given boundary interval is constant. Only while the shockwave travels through the interval does non-trivial time evolution take place.   

Secondly, when at least one of the temperatures $T_L$ and/or $T_R$ is large compared to the interval length $\ell$, the entanglement entropy shows a mostly linear behaviour as a function of time. This is exemplified in figure \ref{fig:UniversalFormula}, where the time dependent entanglement entropy is rescaled as
\begin{equation}
f_A(\rho ) \equiv \frac{S_A(t) - S_A(t=0)}{S_A(t=\infty) - S_A(t=0)} \qquad \textrm{with} \qquad \rho \equiv (t-x_{min})/\Delta t  \,, \label{eq:fA} 
\end{equation}
where $\Delta t =\ell= x_{max} - x_{min}$. This rescaling serves to map the time-dependent part of the curve $S_A(t)$ to the unit square of the axes of figure \ref{fig:UniversalFormula}. Interestingly, such a linear increase is reminiscent to the similar linear increase of entanglement entropy observed in \textit{entanglement tsunamis} \cite{AbajoArrastia:2010yt,Balasubramanian:2010ce,Balasubramanian:2011ur,Balasubramanian:2011at,Hartman:2013qma,Liu:2013iza,Li:2013sia,Liu:2013qca,Leichenauer:2015xra}. However, thirdly, when both $T_L$ and $T_R$ are small compared to the interval length $\ell$, we obtain the universal low-temperature approximation formula \cite{Megias:2015tva,Megias:2016vae,Erdmenger:2017gdk}
\begin{equation}
f_A(\rho ) \simeq 3\rho^2-2\rho^3  \ \ \ \text{for}\ \ \ 0 \le \rho \le 1 \,. \label{eq:fAuniversal}
\end{equation} 

\subsection{Bounds on entropy increase}

The above mentioned similarity between the time dependence of entanglement entropy in entanglement tsunamis and the system investigated in \cite{Bhaseen:2013ypa,Erdmenger:2017gdk} motivates the study of bounds on the entropy increase rate. To this end, we define the averaged entropy increase rate \cite{Erdmenger:2017gdk},
\begin{align}
v_{av}&\equiv\frac{\Delta S}{\Delta t}=\frac{L}{4G\ell}\log\left(\frac{T_L \sinh(\pi\ell T_R)}{T_R \sinh(\pi\ell T_L)}\right),
\end{align}
and the normalised averaged entropy increase rate
\begin{align}
\tilde{v}_{av}&\equiv\frac{v_{av}}{s_{eq}},\ \ 
|\tilde{v}_{av}|\leq\left|\frac{T_R-T_L}{T_R+T_L}\right|\leq1,
\label{averagerate}
\end{align}
where $s_{eq}=\frac{L}{4G}\pi(T_L+T_R)$ is the entropy density of the final steady state. The bound shown in \eqref{averagerate} is exactly the one expected from the study of entanglent tsunamis in $1+1$ dimensional CFTs \cite{AbajoArrastia:2010yt,Balasubramanian:2010ce,Balasubramanian:2011ur,Balasubramanian:2011at,Hartman:2013qma,Liu:2013iza,Li:2013sia,Liu:2013qca,Leichenauer:2015xra}. Furthermore, we can define the normalised \textit{momentary} entropy increase rate
\begin{align}
\tilde{v}\equiv\frac{1}{s_{eq}}\frac{dS(\ell,t)}{dt},
\label{rate}
\end{align}
which in all cases that we studied analytically or numerically also satisfies the bound $|\tilde{v}|\leq1$.

\subsection{Entanglement inequalities}

Another area of interest is the validity of \textit{entanglement inequalities}. It is known that the RT prescription for the calculation of entanglement entropy automatically ensures certain inequalities, amongst others the \textit{strong subadditivity} (SSA) \cite{doi:10.1063/1.1666274,Headrick:2007km}
\begin{align}
&S(AB) + S(BC) - S(ABC) - S(B) \geq 0,
\label{SSA}
\end{align}
where e.g.~$S(BC)$ is the entanglement entropy of the union of the two boundary regions $A$ and $B$. In the time-dependent case where the HRT prescription applies, these inequalities are also satisfied under the assumption that certain energy conditions hold \cite{Wall:2012uf,Prudenziati:2015cva}. Using the RT prescription, several other entanglement inequalities involving $n=5$ or more intervals have been derived in \cite{Bao:2015bfa}, one of them taking the form
\begin{align}
&S(ABC) + S(BCD) + S(CDE) + S(DEA) + S(EAB) - S(ABCDE)
\nonumber
\\
&-S(BC) - S(CD) - S(DE) - S(EA) - S(AB)\geq 0.
\label{Oo1}
\end{align}
The study of such entanglement entropy inequalities in AdS/CFT is of considerable interest for the understanding of holography.  For example, the study of such inequalities for time-dependent setups may provide insights into the role of energy conditions in holography \cite{Lashkari:2014kda}.

\begin{figure}[htb]
		\includegraphics[width=0.55\textwidth]{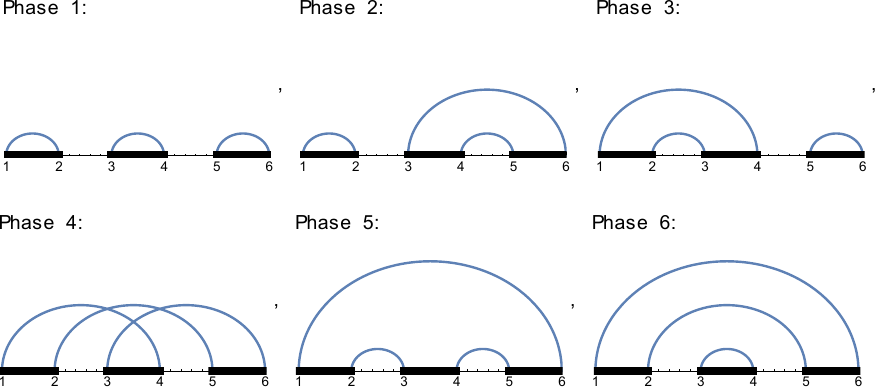}\hspace{0.05\textwidth}%
		\begin{minipage}[b]{0.4\textwidth}	\caption{$6$ possible geodesic configurations for the calculation of entanglement entropy via the HRT proposal for $n=3$ boundary intervals. The 4th configuration is considered to be unphysical. According to the nomenclature of \cite{Ben-Ami:2014gsa}, the phase 6 is referred to as \textit{engulfed} phase.}
			\label{fig::n3case}
		\end{minipage}
	\end{figure}

In \cite{Erdmenger:2017gdk}, we hence numerically tested, in the time dependent model of section \ref{sec::intro}, the validity of \eqref{Oo1} and several other entanglement entropy inequalities proven or conjectured to hold for static cases . In order to do so, we had to take into account the different geodesic configurations or \textit{phases} that may play a role in the RT or HRT prescription for the union of $n\geq2$ boundary regions. See e.g.~figure \ref{fig::n3case} for the case $n=3$. Our results can be summarised as follows: For $n=3$ intervals $A$, $B$ and $C$, strong subadditivity \eqref{SSA} is satisfied, as expected based on \cite{Wall:2012uf,Prudenziati:2015cva}. In contrast to the situations investigated in \cite{Ben-Ami:2014gsa}, even the \textit{engulfed phase} (see figure \ref{fig::n3case}) can be the relevant one in specific examples. Generically, this can happen when the middle interval is very small compared to the gap between the other two intervals. 
	
For $n=4$ intervals $A$, $B$, $C$ and $D$, the only inequality that we are checking is the \textit{positivity of four-partite information} 
\begin{align}
S(A)+S(B)+...-S(AB)-S(AC)-...+S(ABC)+S(ABC)+...-S(ABCD)\geq 0,
\label{I4}
\end{align}
which was conjectured to hold in \cite{Alishahiha:2014jxa,Mirabi:2016elb}. We, however, find a number of examples for sets of four intervals where this inequality is violated. As it was pointed out in \cite{Hayden:2011ag} and was explicitly checked by us, this happens already in holographic systems with static bulk-spacetime duals.  
	
\begin{figure}[htb]
	\includegraphics[width=0.55\textwidth]{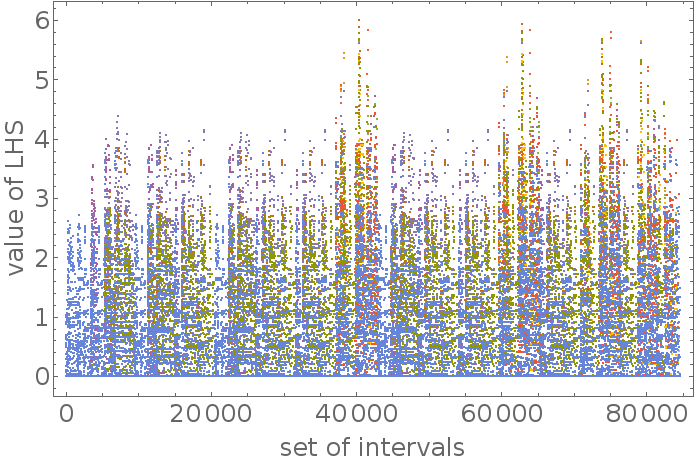}\hspace{0.05\textwidth}%
	\begin{minipage}[b]{0.4\textwidth}\caption{The left-hand sides of the inequality \eqref{Oo1} for $84579$ different sets of five intervals for which this quantity does not vanish trivially. Clearly there are no violations of the inequality \eqref{Oo1}. The different colors in the figure stand for different permutations of the inequality, obtained from \eqref{Oo1} by exchanging the labels ($A$, $B$, ...) of the intervals. We are here displaying results for the example where $T_L=9$, $T_R=1$ and boundary time $t=1$.       
		}
		\label{fig::Oo1}
	\end{minipage}
\end{figure}

For $n=5$ intervals $A$, $B$, $C$, $D$ and $E$, we find numerous violations of the \textit{negativity of five-partite information} \cite{Alishahiha:2014jxa,Mirabi:2016elb}, see the similar discussion for $n=4$. Furthermore, we check the inequality \eqref{Oo1} as well as the other inequalities for $n=5$ intervals derived in \cite{Bao:2015bfa}. The result is that we find \textit{not a single case} in which any of these inequalities is violated, see for example figure \ref{fig::Oo1}. We view this as a clear indication that these inequalities (such as \eqref{Oo1}), although so far only proven in the static case, will generally also hold in physical time-dependent cases.

\ack

MF was supported by NCN grant 2012/06/A/ST2/00396. EM is supported by the U. Pa\'{\i}s Vasco UPV/EHU, Spain, as a Visiting Professor. We would also like to thank the organisers of the Karl Schwarzschild meeting 2017 for their hospitality.


\section*{References}


\providecommand{\newblock}{}

\end{document}